\documentclass[11pt,twoside]{article}
\usepackage{macroaaa-eng}
\usepackage{psfig}
\usepackage{graphicx}

\usepackage[T1]{fontenc} 

\usepackage{latexsym}
\usepackage{verbatim}

\begin{document}

\vskip 1.0cm \markboth{I.~Sidelnik et. al.}{First buried muon
counter prototype for the Auger Observatory}
\pagestyle{myheadings}

\vspace*{0.5cm}
\title{First buried muon counter prototype for the Auger Observatory}
\author{I. P. Sidelnik$^{1}$,  B. Wundheiler$^{1}$, E. Colombo$^{1}$, A. Etchegoyen$^{1}$,
A. Ferrero$^{1}$, M. Platino$^{1}$, O. Wainberg $^{1}$ }

\affil{ $^1$ Centro At\'omico Constituyentes (Comisi\'on Nacional de Energ\'ia At\'omica/CONICET/UTN-FRBA)}

\begin{abstract}
AMIGA (Auger Muons and Infill for the Ground Array) constitutes an
enhancement for the Pierre Auger Observatory. It consists of a
denser array of surface detectors and muon counters whose
objective is both to extend the detection range down to 10$^{17}$
eV and to help towards mass composition determination. The latter
is to be achieved with muon counters since the shower muon content
is one of the best parameter for particle type identification. In
this work, we present the study of a muon counter prototype. The
prototype was buried 3 m deep in an effort to avoid signal
contamination from the shower electromagnetic component. We study
the performance of the detector before and after burying it with
its associated electronic components. The detector validation is
performed from signal analysis of charged particles traversing the
counter.
\end{abstract}

\section{Introduction}

The cosmic ray energy spectrum presents four main features (Nagano \& 
Watson, 2000) which give rise to abrupt spectral index
changes: the knee, the second knee, the ankle and the GZK cutoff.
The knee is at $\sim4\times10^{15}$ eV, the second knee at $\sim
4\times10^{17}$ eV, the ankle at $\sim3\times10^{18}$, and 
finally the GZK cutoff, i.e. the suppression of the cosmic ray
flux at energies above $\sim 4 \times10^{19}$ eV. The Auger
collaboration has recently shown experimental evidence of the
ankle (Tokonatsu, 2007), the GZK-cutoff (Abraham et al., 2008a),
and of the anisotropy in the arrival direction of
the cosmic rays with energies above $\sim6\times10^{19}$ eV
(Abraham et al., 2008a) which is a further confirmation of the flux cutoff.

The ankle region may actually be considered as a dip raging from
the second knee up to $\sim 10^{19}$ eV and in an attempt to
fully study these two spectrum traits Auger is building two
detector enhancements: AMIGA (Etchegoyen, 2007) and HEAT
(High Elevation Auger Telescopes) (Klages, 2007). The
physical interpretations of the second knee and ankle are not yet
clear but still they are closely linked. They are assumed to be
related to the transition from galactic to extra-galactic cosmic
ray sources and from dominant heavy to dominant light primary
compositions (Allard et al., 2005, Berezinsky et al., 2004, Wibig
and Wofendale, 2005).

The Pierre Auger Observatory  (Abraham et al., 2004) is built to
detect cosmic rays of the highest energy
with two distinct design features, a large size
and a hybrid detection system in an effort to detect a large
number of events per year with reduced systematic uncertainties.
It will have a southern and a northern component. The former is
located in the west of Argentina, in the Province of Mendoza where
it spans an area of 3000 km$^{2}$ covered with a surface detector
(SD) system of 1600 water Cherenkov detectors (Allekote et al., 2008)
deployed on a 1500 m triangular grid plus four buildings on the
array periphery lodging six fluorescence detector (FD) telescopes
each one with a $30^{\circ }\times 30^{\circ}$ elevation and
azimuth field of view. As above mentioned it now encompasses two
enhancements and in this work we are going to concentrate on the
AMIGA muon counters.

\section{AMIGA muon counters \label{sec:AMIGA}}

\begin{figure}[htp] 
\begin{center}
\includegraphics[width=7 cm]{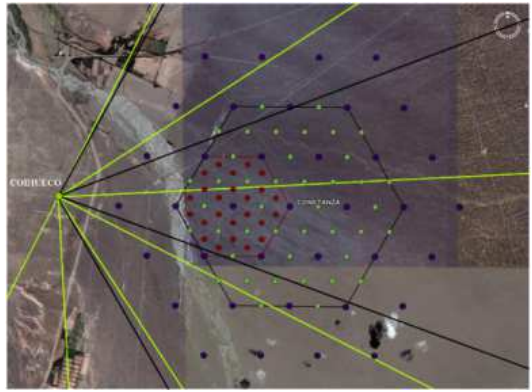}
\caption{Auger enhancements layout. Light and dark lines limit the
$0^{\circ}-30^{\circ}\times 0^{\circ}-30^{\circ}$ and
$30^{\circ}-60^{\circ}\times 0^{\circ}-30^{\circ}$ elevation
and azimuth field of view for the original 6 fluorescence
telescopes and the HEAT 3 telescopes on Cerro Coihueco,
respectively. The two hexagons limit the AMIGA areas of 5.9 and
23.5 km$^{2}$ with 433 and 750 m triangular grid detector
spacings, respectively. Each dot within these hexagons represents
a pair of a water Cherenkov tank and a muon counter. The center
dot is the Constanza pair placed $\sim$6.0 km away from Cerro
Coihueco. \label{AMIGA-map}}
\end{center}
\end{figure}

The Observatory was original designed to be fully efficient at
$\sim 3\times 10^{18}$ eV for the surface array and at $\sim
10^{18}$ eV for the hybrid mode. The enhancements (see Fig. \ref{AMIGA-map})
will extend these efficiencies down to $10^{17}$ eV
and $\sim 2 \times 10^{17}$ eV, respectively. Decreasing energies
imply both a flux increment and a smaller shower lateral
distribution spread and therefore AMIGA will encompass a smaller
area with a denser array (Etchegoyen, 2007).

The muon number and $X_{max}$ (the position where the airshower
attains its maximum development) are the best two indicators of
the chemical composition of cosmic rays. Auger measures $X_{max}$
using its fluorescence telescopes and it will measure the shower
muon content (Supanitsky et al., 2008) with plastic scintillators. To
achieve this objective, along with each SD of the graded denser
array, a 30 m$^{2}$ area muon counter will be buried in order to
avoid electromagnetic contamination.. A first 5 m$^{2}$ module will
be buried 2.25 $\pm$ 0.05 m underground by the end of 2009.
The scintillator modules are not supposed to
reconstruct the shower parameters (which will be performed by the
FD and SD systems) but just to count muons.

The scintillator strips were developed following the MINOS design
(MINOS, 1998) at the ``Scintillator Fabrication Facility'' at
the Fermi National Accelerator Laboratory (FNAL). Seven counters
will be burried by the beginning of 2011 and each of these counters
will have four modules with 4.1 cm wide $\times$ 1.0 cm high
strips, two of them 400 cm long and the other two 200 cm long, in
order to check pileup close to the shower core. They are made of
extruded polystyrene doped with fluors and co-extruded with a
TiO$_{2}$ reflecting coating with a groove in where a wave length
shifter (WLS) fiber is glued and covered with a reflective foil
(Etchegoyen, 2007). Each module will consist of 64 strips
with the fibers ending in an optical connector matched to a 64
multianode photo multiplier tube (PMT) from the Hamamatsu ultra
bialkali H8804-200MOD series (2 mm $\times$ 2 mm pixel size). The
module enclosure will be made of PVC.

Before proceeding to install muon counters at the field in the
Observatory site, it is convenient to bury and deploy a prototype
in a field close to a laboratory environment within which the
process can be conveniently validated under emulated AMIGA working
conditions. The main objective of this work is to report on such a
buried muon counter prototype.

\section{Main objective \& Procedure}

\begin{figure}  
\begin{center}
\hbox{
   \raisebox{15pt}{\psfig{figure=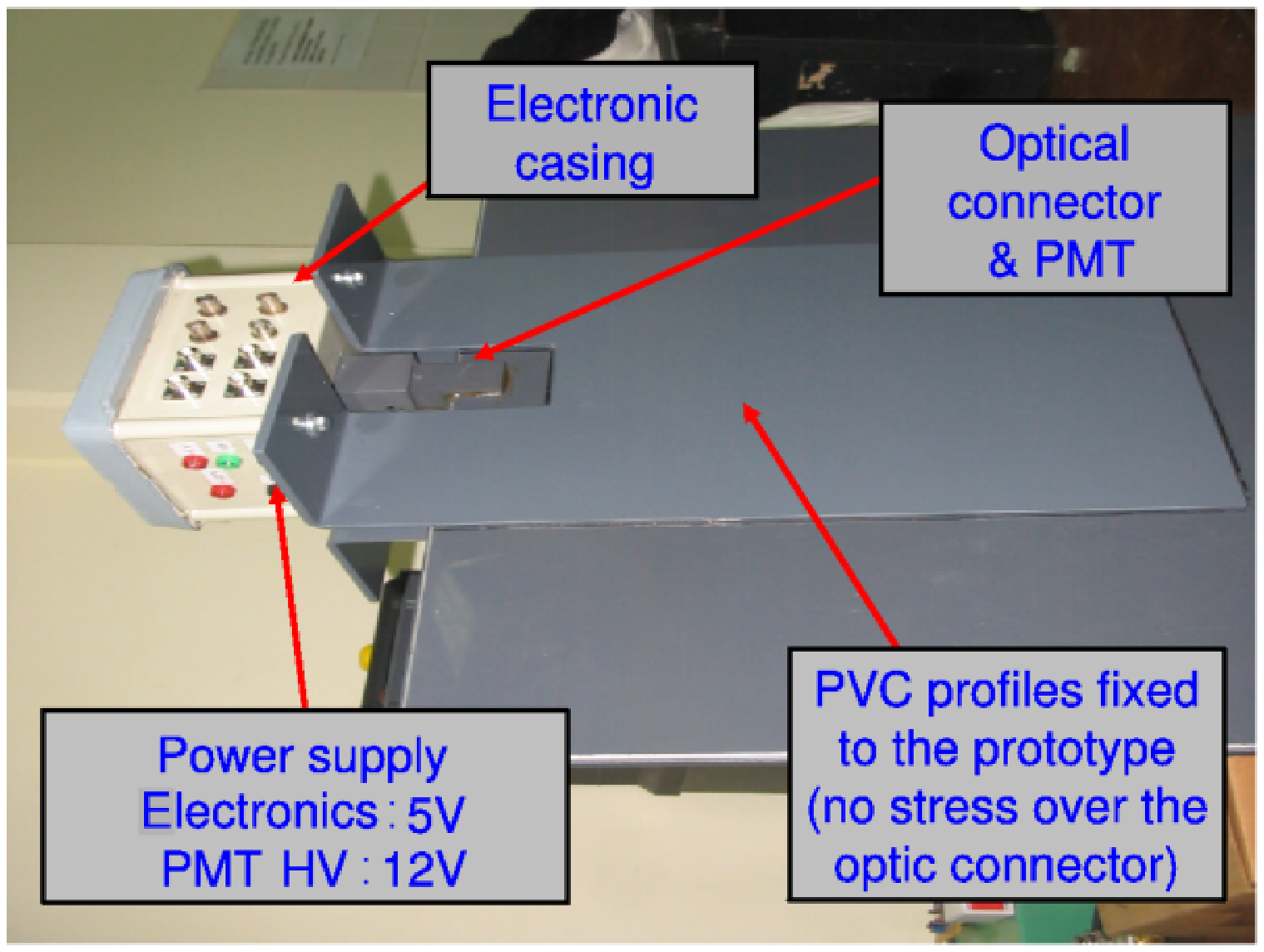,width=6cm}}
   \psfig{figure=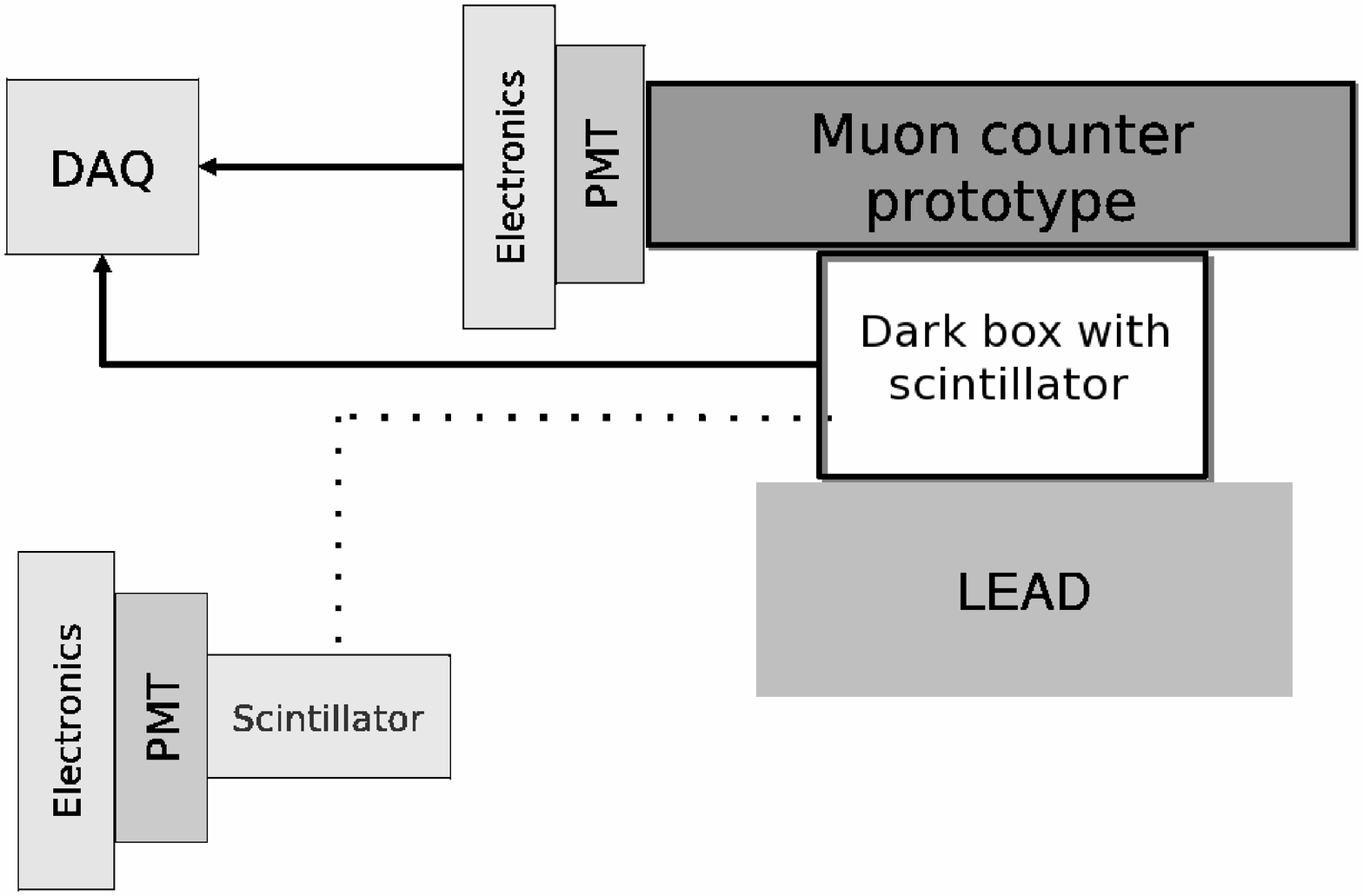,width=6cm}
        }
\caption{\label{fig:Muon-counter-main}Muon counter main features
and schematics of the experimental setup. {\it lhs}: Detail of the
electronic underground enclosure. {\it rhs}: Coincidence box used
as an external trigger, dash line points to experimental setup
inside the box (not in scale).}
\end{center}
\end{figure}

A prototype was developed for AMIGA by Argonne National Laboratory (ANL)
and FNAL. This module is 135 cm long, 75 cm wide (16 strips), and 1 cm
thick with 1.2 mm WLS Kuraray green optical fiber glued in each
strip groove. A 16 multianode Hamamatsu H8711-06 PMT was used
with an 8 channel associated electronics board in order to acquire
the PMT analogic signals.

To protect the electronic board and connections from humidity and
corrosion, an ad-hoc enclosure box capable to endure buried
conditions was designed and assembled. Figure
\ref{fig:Muon-counter-main} ({\it lhs}) shows details of the electronic
enclosure with output (top wall) and power supply (side wall)
connectors. The box is connected to the PMT which in turns links
to the module via the optical connector to which the WLS fibers
are glued. Data acquisition was performed with a Tektronix TDS
3032 digital oscilloscope (300 MHz of band width and 2.5 GS/s of
sample rate).

The experiment was set to measure quasi-vertical impinging
atmospheric muons and to ensure this, a time coincidence was
requested. This coincidence was attained by adding to the
experimental setup a short scintillator strip connected to a PMT
inside a dark box. Its signals were channelled to the oscilloscope
external trigger (see Figure \ref{fig:Muon-counter-main} ({\it rhs})).

\begin{figure}  
\hbox{
   {\psfig{figure=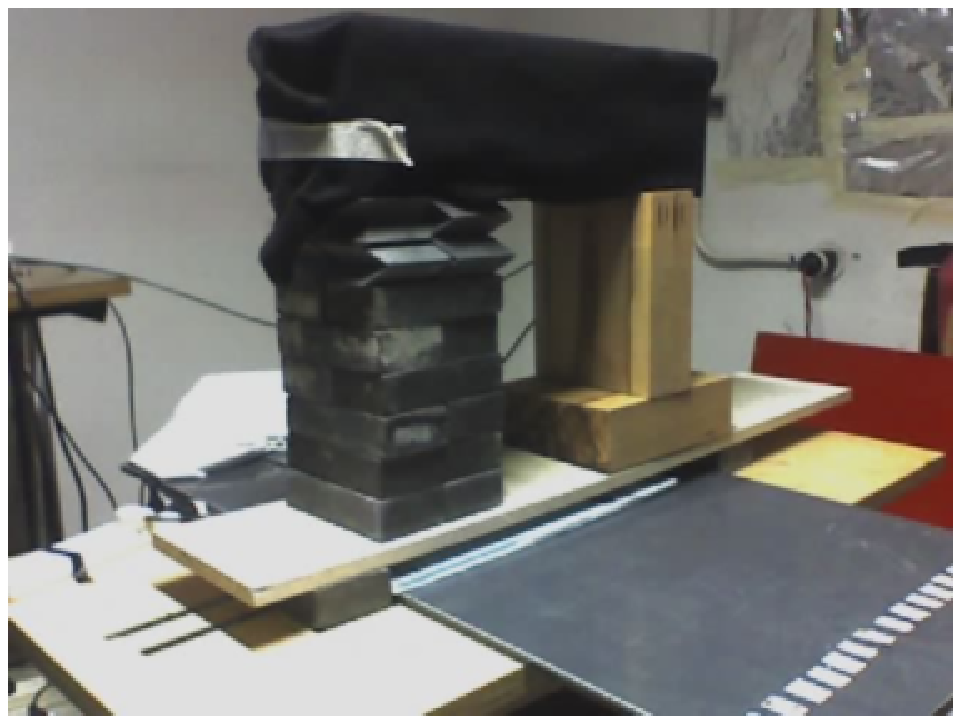,width=6.5cm}}
          \hspace*{0.3cm}
   \psfig{figure=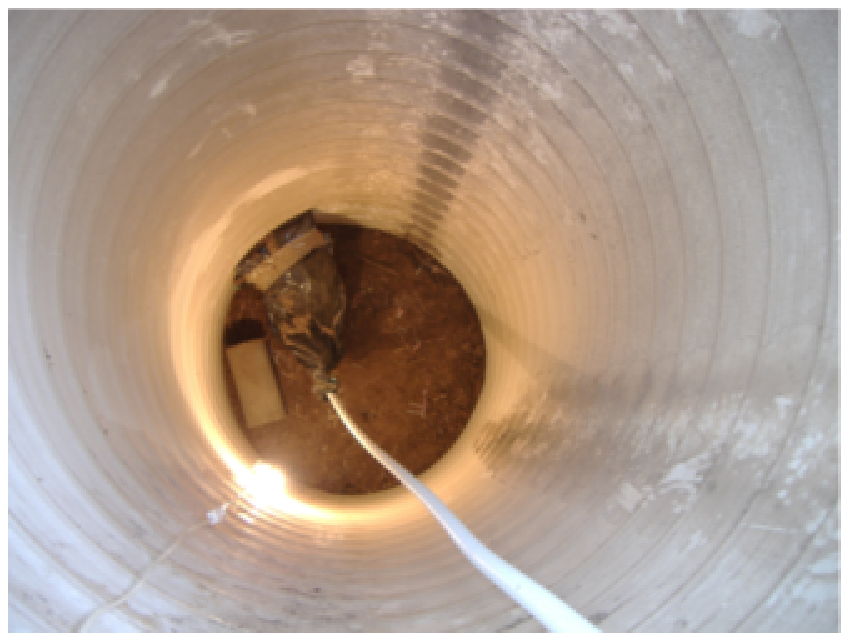,width=6.5cm}
        }
\caption{\label{fig:Set-up-A}Setup A \& B: laboratory conditions
and buried detector. {\it lhs:} suspended lead column on the
module prototype in the laboratory, {\it rhs:} service pipe to
gain access to the electronics.}
\end{figure}

With the purpose of understanding the effect of the soil over the
muon counter  we carried out two different experiments, with
the counter set in the laboratory (setup A) and afterwards with
the same counter buried in soil (setup B).

In setup A we set off by subjecting the prototype under stress for
a period of a month by placing it underneath a 45 cm lead column
in order to simulate the equivalent 540 g/cm$^{2}$ soil pressure (3 m deep
with a soil density of 1.8 g/cm$^{3}$). Experiments
were performed after this test in an effort to discard any module
mechanical damage prior to burring it, stress damages over longer
period of time were to be tested after burying the module. The
lead stress was applied at the module center over an area four
strip wide $\times$ 30 cm long. Also, and in order to avoid ground
electromagnetic contamination, a lead shielding under the detector
was used.

A test comparison was performed under the same shielding (i.e.
amount of lead) causing and no causing stress, i.e. directly
resting on the module or suspended just above it (see Figure
\ref{fig:Set-up-A} ({\it lhs})). Results will be discussed in Sec. 4.

In setup B we repeated these conditions in a 3 m well excavated
near the Tandar Laboratory, first making the measurements without
dirt and then covering the detector with it. For a month the
counter was kept buried during this stage of the experiment.

\begin{figure}[!htp]  
\hbox{\hspace{1.0cm}
  \psfig{figure=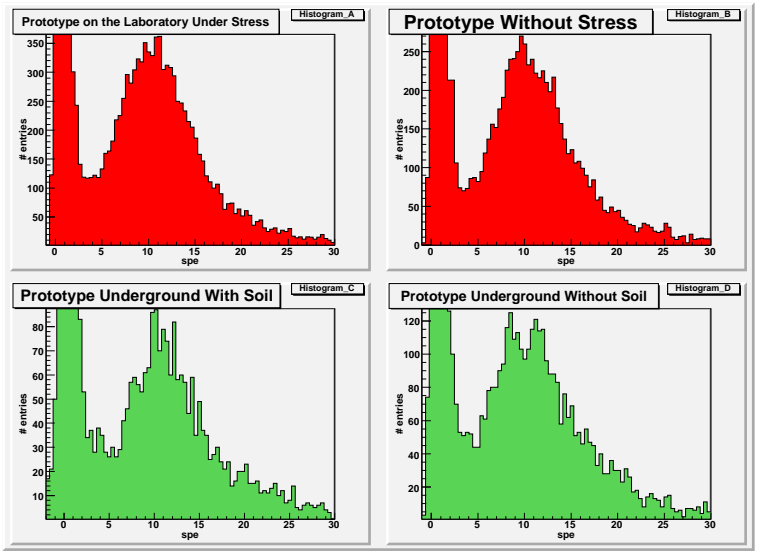,width=10cm}
        }
\caption{\label{fig:Charge}Charge Histograms as a result of the
measurements made with setup A \& B: comparison between results
show similar features in all cases. {\it Top:} laboratory setup
tested with and without stress. {\it Bottom:} buried prototype
with and without dirt.}
\end{figure}

As a result of these different setups we had four different
experimental conditions: lead suspended above and resting on the
prototype, and underground with and without soil. On the {\it lhs} of
Figure \ref{fig:Set-up-A} we show the experimental setup mounted
in the laboratory with a column of lead and the coincidence box,
while on the {\it rhs} we show a detail of the service pipe (3 m tall
1 m diameter) installed to gain access to the buried electronics.

\section{Results}\label{sec:results}
Total charge histograms were performed integrating the pulses generated by
each traversing muon. In order to study and compare different scenarios
the histograms were normalized to the single photoelectron (spe)
charge so the comparison can be made in terms of the spe for the four setups.

Figure \ref{fig:Charge} shows the normalized charge histograms for
the prototype for each experiment layout, using one representative 
pixel of the PMT used. Fitting the zone associated with muon signals,
we found a mean value of 10.55 $\pm$ 2.19 pe for the four setups,
in agreement with our previous results (Krieger, 2008).
The analyses show that the response of the buried prototype
is the same as in laboratory conditions. So we conclude that the
muon signal provided by our counter appears not been affected by
neither applying stress or burying it in a 3 m well covered with
soil, supporting the baseline AMIGA muon counters as a tool to
establish the muon content of an extended air shower.

\section{Conclusions}

A new design and construction of a prototype electronic enclosure and
its assembly onto the muon counter case was developed. Of more relevance
was the design, construction and operation of the service pipe which will
be used in AMIGA.

We made the first positive experiment in working conditions testing
a muon counter and its electronics. We showed that burying the detector
3 m deep does not affect the measurements.

We also showed muon histograms do not change in either laboratory
(with and without stress) or buried conditions.

\end{document}